\documentclass[twocolumn,prb,showpacs,preprintnumbers,amsmath,amssymb]{revtex4}
\usepackage{amsmath,amssymb,color,graphicx,epsfig}
\usepackage{wrapfig}

\usepackage{amssymb,amsfonts,amsmath,color,fontenc}
\usepackage{rotating,epsfig}

\newcommand{\trans}{\scriptscriptstyle\mskip-1mu\top\mskip-2mu}

\newcommand{\calC}{\mathcal{C}}
\newcommand{\calD}{\mathcal{D}}
\newcommand{\calS}{\mathcal{S}}
\newcommand{\calR}{\mathcal{R}}

\newcommand{\dro}{\,\text{d}\rho}
\newcommand{\dtheta}{\,\text{d}\theta}

\newcommand{\half}{{\textstyle{\frac{1}{2}}}}

\newcommand{\bfe}{\boldsymbol{e}}
\newcommand{\bfx}{\boldsymbol{x}}

\newcommand{\bfxi}{\boldsymbol{\xi}}

\newcommand{\xir}{\bfxi_\rho}

\newcommand{\xit}{\bfxi_{\theta}}

\newcommand{\er}{\bfe_{r}}
\newcommand{\et}{\bfe_{\theta}}
\newcommand{\ez}{\bfe_{z}}

\newcommand{\ti}{\times}

\newfont{\tenbss}{bbmss12}

\newcommand{\calF}{\mathcal{F}}
\newcommand{\calE}{\mathcal{E}}
\newcommand{\calH}{\mathcal{H}}

\newfont{\tenbfsl}{cmbxti12}

\newcommand{\be}{\begin{equation}}
\newcommand{\bea}{\begin{eqnarray}}
\newcommand{\ee}{\end{equation}}
\newcommand{\eea}{\end{eqnarray}}

\begin{document}

\title{Stability of discoidal high-density lipoprotein particles}

\author{Mohsen Maleki and Eliot Fried                   
\\[8pt]
\small Department of Mechanical Engineering,
McGill University,
Montr\'eal, QC H3A 0C3, Canada
}

\begin{abstract}
Motivated by experimental and numerical studies revealing that discoidal high-density lipoprotein (HDL) particles may adopt flat elliptical and nonplanar saddle-like configurations, it is hypothesized that these might represent stabilized configurations of initially unstable flat circular particles. A variational description is developed to explore the stability of a flat circular discoidal HDL particle. While the lipid bilayer is modeled as two-dimensional fluid film endowed with surface tension and bending elasticity, the apoA-I belt is modeled as one-dimensional inextensible twist-free chain endowed with bending elasticity. Stability is investigated using the second variation of the underlying energy functional. Various planar and nonplanar instability modes are predicted and corresponding nondimensional critical values of salient dimensionless parameters are obtained. The results predict that the first planar and nonplanar unstable modes occur due to in-plane elliptical and transverse saddle-like perturbations. Based on available data, detailed stability diagrams indicate the range of input parameters for which a flat circular discoidal HDL particle is linearly stable or unstable.
\end{abstract}


\maketitle

\section{Introduction}

The packaging and transport of water-insoluble cholesterol in the bloodstream are mediated by lipoprotein particles. In ``reverse cholesterol transport,'' high-density lipoprotein (HDL) particles scavenge cholesterol from tissues and other {types of lipoprotein particles} and deliver it to the liver for excretion into bile or other use. A comprehensive understanding of the biophysical basis for the vasculoprotective functionalities of HDL particles {is} essential to developing effective strategies to prevent, diagnose, and treat atherosclerosis. However, as Vuorela et al.~\cite{vuorela10} observe: ``\emph{The functionality of HDL has remained elusive, and even its structure is not well understood.}"

During reverse cholesterol transport, an HDL particle sustains shape transitions that are accompanied by changes in the conformation of its apolipoprotein building block apoA-I. Davidson \& Silva~\cite{davidson05} explain that the functionality of apoA-I is linked to its conformational variations and emphasize the need to understand the diverse range of conformations that it adopts in its lipid-free and lipid-bound forms. A discoidal HDL particle consists of a lipid bilayer {bound} by an apoA-I chain. Camont et al.~\cite{camont11} argue that the low lipid content and high surface fluidity of discoidal HDL particles induces conformational changes of apoA-I that result in enhanced exposure to its aqueous surroundings and, thus, in an increased capacity to acquire blood lipids. Using all-atom molecular dynamics (MD) simulations, Catte et al.~\cite{catte06} predict that assembling a flat circular HDL particle from a lipid-free apoA-I chain involves the formation of intermediate nonplanar, twisted, saddle-like particles. Coarse-grained molecular dynamics simulations of Shih et al.~\cite{shih07a,shih07b} and experiments of Silva et al.~\cite{silva05}, Miyazaki et al.~\cite{miyazaki09}, and Huang et al.~\cite{huang11} confirm this prediction. {In addition, experimental results of Skar-Gislinge et al.~\cite{Skar-Gislinge} reveal that HDL particles exhibit an intrinsic tendency to adopt planar, elliptical configurations.}

MD simulations have provided valuable insight regarding the molecular interactions that govern the assembly and dynamics of discoidal HDL particles. However, the small time steps needed to correctly capture the highest frequency {of} molecular vibrations and preserve numerical accuracy make it difficult to access time scales long enough to determine equilibria or draw conclusions regarding stability. For these purposes, continuum models provide a valuable complement to MD simulations. In particular, continuum models have been used with remarkable success to determine equilibria and study stability in biomembranes and biomolecules.

Inspired by the aforementioned experimental and MD simulations, a continuum mechanical model for the equilibrium and stability of a flat circular HDL particle is presented. Guided by  prevalent continuum models of biomembranes and biomolecules, the bilayer is treated as a two-dimensional fluid film endowed with surface tension and resistance to bending and the apoA-I chain as a one-dimensional inextensible, twist-free, elastic filament endowed with resistance to bending. The bilayer and apoA-I chain are required to be perfectly bonded, in which case the boundary of the fluid film and the elastic filament must have the same shape. A variational description of the equilibrium of a discoidal HDL particle is provided. A flat, circular shape is chosen as a reference configuration. To study the linear stability of the reference shape, infinitesimal perturbations involving both planar and transverse components are considered. Such perturbations can be caused by thermal fluctuations of the lipid bilayer or the apoA-I chain or by interactions between the HDL particle and its environment. Closed-form analytical solutions for the linearized equilibrium conditions are obtained and stability is explored via the second-variation condition. In addition, available values of the physical parameters that enter the model are used to determine the range of inputs under which a flat, circular HDL particle is linearly stable or unstable. Lastly, connections between our result and previous experimental measurements and numerical simulations are made.

\section{Energetics of a discoidal HDL particle}
\label{re}
Geometrically, a discoidal HDL particle is treated as a smooth, orientable surface $\calS$ with boundary $\calC=\partial\calS$. The interior and boundary of $\calS$ correspond, respectively, to the bilayer and apoA-I components of the particle. Following convention, $H$ and $K$ denote the mean and Gaussian curvature of $\calS$ and $\kappa$ denotes the curvature of $\calC$. 

To capture the energetics of the bilayer, $\calS$ is endowed with a uniform surface tension $\sigma$ and an areal bending-energy density
\be
\psi=\half\mu H^2+\bar\mu K,
\label{e1}
\ee
of the type put forth by Canham~\cite{Canham} and Helfrich~\cite{Helfrich}, where $\mu>0$ and $\bar{\mu}$ are the splay and saddle-splay moduli. The relevance of spontaneous curvature, which  ordinarily appears in the Canham--Helfrich model, to discoidal HDL particles has yet to be investigated and, thus, is omitted from \eqref{e1}.

To capture the energetics of the apoA-I chain, $\calC$ is endowed with a lineal bending-energy density $\varphi$ depending on the curvature $\kappa$ of $\calC$ and its arclength derivative $\kappa'$. The latter dependence is included to account for the energetic cost {of }large, localized curvature variations associated with kinks on the apolipoprotein chain discussed by Brouillette et al.~\cite{Brouillette} and Klon et al.~\cite{Klon}. For simplicity, it is assumed that 
\begin{align}
\varphi
=\half\alpha\kappa^2+\half\beta(\kappa')^2,
\label{GF22}
\end{align}
where $\alpha>0$ is the constant flexural rigidity of $\calC$ and $\beta\ge0$ is a higher-order generalization thereof. Since $\calC$ is closed, including a quadratic coupling term proportional to $2\kappa\kappa'=(\kappa^2)'$ in $\varphi$ would not alter the net potential energy and no generality is lost by neglecting such a contribution. The particular choice \eqref{GF22} of $\varphi$ is a special case of a general expression for the lineal free-energy density of a polymer chain proposed by Zhang et al.~\cite{Zhang}, who allow for arbitrary dependence on $\kappa$, $\kappa'$, and the torsion $\tau$ of $\calC$.
Granted the foregoing assumption and that external forces associated with gravity, van der Waals interactions, or flow-related forces are negligible, the net potential-energy of a discoidal HDL particle is given by
\be
\calE=\int_{\calS}(\sigma+\psi)+\int_{\calC}\varphi.
\label{GF0}
\ee
As a surface with boundary, $\calS$ has Euler characteristic equal to unity. On using \eqref{e1} and \eqref{GF22} in \eqref{GF0} and applying the Gauss--Bonnet theorem (taking into consideration that $\calC$ is assumed to be smooth), the net potential-energy $\calE$ becomes
\be
\calE=\calE_a+\calE_l+2\pi \bar \mu,
\label{GF1}
\ee
where
\be
\calE_a=\int_{\calS}(\sigma+\half\mu H^2)
\label{Ea}
\ee
and
\be
\calE_l=\int_{\calC}\big(\half\alpha\kappa^2+\half\beta(\kappa')^2-\bar{\mu}\kappa_g\big),
\label{El}
\ee
denote the effective areal and lineal potential energies, with $\kappa_g$ being the geodesic curvature of $\calC$. Without loss of generality, the additive constant $2\pi\bar\mu$  in \eqref{GF1} is disregarded hereafter.

The assumed inextensibility of the apoA-I chain is imposed by working with the augmented net potential-energy
\be
\calF=\calE_a+\calE_l+\int_{\calC} \lambda,
\label{func}
\ee
where $\lambda$ is an unknown Lagrange multiplier.
\section{Parameterization and nondimensionization}
Let $\calD=\{(r,\theta)\in\mathbb{R}^2:0\leq r\leq R,0\leq\theta\leq2\pi\}$ denote the disk of radius $R$. The surface and boundary of a discoidal HDL particle can then be described by a smooth function
\be
\bfx:\calD\rightarrow\mathbb{R}^3.
\ee
Due to the inextensibility of $\calC$, $\bfx$ must satisfy
\be
|\bfx_\theta(R,\theta)|=R,
\quad
0\le\theta\le2\pi.
\label{xinextensible}
\ee
With this choice, the bilayer and apoA-I chain are represented by 
\be
\bfx(r,\theta),
\quad
0\le r<R,
\quad
0\leq\theta\leq2\pi,
\label{bilayer}
\ee
and
\be
\bfx(R,\theta),
\quad
0\leq\theta\leq2\pi.
\label{chain}
\ee
On determining expressions for the geometrical objects $H$, $\kappa$, $\kappa'$, and $\kappa_g$ consistent with the parametrization {\eqref{bilayer}--\eqref{chain}}, the augmented net potential-energy $\calF$ defined in \eqref{func} can be expressed as a functional of $\bfx$.

It is convenient to present results in dimensionless form via the change of variables
\be
\bfx(r,\theta)=R\mskip1.5mu\bfxi(\rho,\theta),
\quad
r=R\mskip1mu\rho,
\label{nd1}
\ee
in which case the dimensionless reference domain is a disk of radius unity denoted by $\calR$. In addition, it is convenient to introduce the following group of dimensionless quantities
\begin{multline}
\big(\calH,\eta,\bar\eta,\nu,\iota,\epsilon\big)
\\
:=\Big(\frac{{\cal{F}} R}{\alpha},\frac{\mu R}{\alpha},\frac{\bar \mu R}{\alpha}, \frac{\sigma R^3}{\alpha}, \frac{\lambda R^2}{\alpha}, \frac{\beta}{\alpha R^2}\Big).
\label{nd2}
\end{multline}
%
In particular, the dimensionless counterpart $\calH$ of the augmented net potential energy defined \eqref{func} takes the form
\begin{multline}
\calH=\int^{2\pi}_{0}\int^{1}_{0}(\nu+\half\eta R^2H^2)|\xir\ti\xit|\dro \dtheta 
\\
+\int^{2\pi}_{0}(\half R^2\kappa^2+\half\epsilon R^4(\kappa')^2-\bar\eta R\mskip1mu\kappa_g+\iota\big)|\xit|_{\rho=1}\dtheta.
\label{par51}
\end{multline}
For brevity, the adjective `dimensionless' is dropped hereafter.
 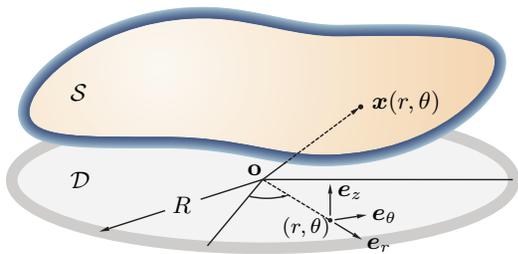
\begin{figure} [t]
  \centering
  \includegraphics [height=1.3in] {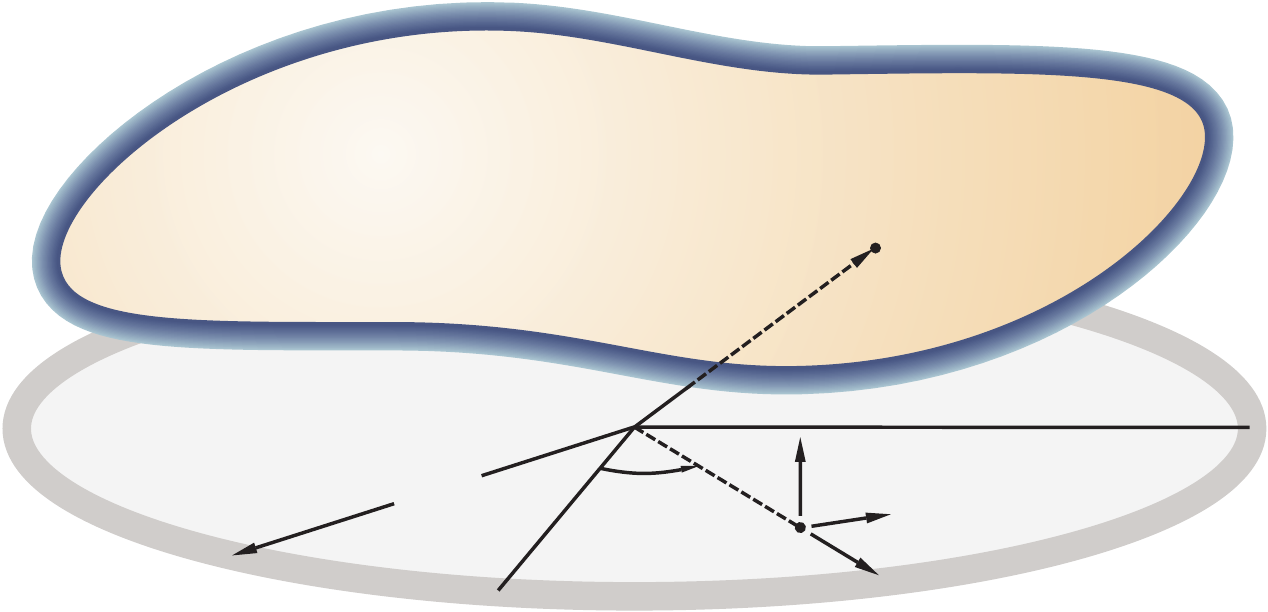}
  \put(-170,58){$\calS$}   
  \put(-170,25){$\calD$}
  \put(-103,30){$\bf{o}$}
  \put(-56,55){$\bfx(r,\theta)$}
  \put(-58,2){${\er}$}
  \put(-56,13){${\et}$}
  \put(-69,22){${\ez}$}
  \put(-90,8){\footnotesize${(r,\theta)}$}
  \put(-131.5,15.5){$R$}
  \\[2pt]
\caption{Schematic of a slightly perturbed discoidal HDL particle and its flat circular reference configuration (in grays). The transverse deformation is exaggerated for  illustrative purposes.}
\label{f1}
\end{figure}
To obtain linearized equilibrium equations and {study the} stability of a discoidal HDL particle, it suffices to use {an} infinitesimal displacement approximation in which the position of a generic point on $\calS$ is given by \eqref{nd1}, with
\be
\bfxi(\rho,\theta)={\bf{o}}+(\rho+u(\rho,\theta))\er+ v(\rho,\theta)\et
+w(\rho,\theta)\ez,
\label{cb1}
\ee
where $\bf{o}$ indicates the origin of the reference disk and $u$, $v$, and $w$ are the components of $\bfxi$ in $\er$, $\et$, and $\ez$ directions, respectively (Fig.\ 1). The inextensibility condition \eqref{xinextensible} becomes
\be
|\xit(1,\theta)|=1,
\quad
0\le\theta\le2\pi.
\label{constr}
\ee

The expansion \eqref{cb1} can be used to express \eqref{par51} componentwise. {The linearized equilibrium conditions arise on expanding all terms in \eqref{par51} up to the second order of $u$, $v$, $w$, including their partial derivatives.} For brevity, the intermediate {calculations} are suppressed.

\section{Equilibrium conditions}
\label{eebc}
 
At equilibrium, the first variation $\dot \calH$ of the functional $\calH$ in \eqref{par51} vanishes. Notice that a superposed dot indicates the first variation. Imposing the requirement $\dot\calH=0$ yields the partial-differential equation
\be
\Delta^2w-\zeta^2 \Delta w=0
\qquad {\rm {on} \ \calR},
\label{sd10}
\ee
with $\Delta$ the Laplacian on $\calR$ and $\zeta=2\sqrt{\nu/\eta}$, and boundary conditions
\begin{multline}
\Big[(\iota+\nu-\half)+\nu(u+v_\theta)+\half(2u+5u_{\theta\theta}+2u_{\theta\theta\theta\theta}
-v_{\theta})
\\
-\epsilon\big(u_{\theta\theta}+2u_{\theta\theta\theta\theta}+u_{\theta\theta\theta\theta\theta\theta}\big)-\big(\iota(u_{\theta}-v)\big)_\theta\Big]_{\rho=1}=0,
\label{sd11}
\end{multline}  
\be
\big[\iota_\theta+(\iota+\nu-\half)(u_\theta-v) \big]_{\rho=1}=0,
\label{sd12}
\ee  
\begin{multline}
\Big[ -\frac{\eta}{4}(\Delta w)_\rho +\nu w_\rho
+\half(3w_{\theta\theta}+2w_{\theta\theta\theta\theta})
\\
-\bar\eta(w_{\theta\theta}-w_{\rho\theta\theta})-(\iota w_\theta)_\theta \Big]_{\rho=1}=0,
\label{sd13}
\end{multline}  
\begin{equation}
\Big[ \frac{\eta}{4} \Delta w +\bar \eta (w_\rho+w_{\theta\theta}) \Big]_{\rho=1}=0.
\label{sd14}
\end{equation}  
%
%
Moreover, the linearized version of the inextensibility condition \eqref{constr} requires that
\be
(u+v_\theta)_{\rho=1}=0.
\label{sdinex}
\ee

The equilibrium condition \eqref{sd10}, which governs the local geometry of the lipid bilayer, is the linearized version of the shape equation familiar from works on vesicles. The boundary conditions \eqref{sd11}{, \eqref{sd12}, and} \eqref{sd13} express force balance on $\calC$ in the $\er$, $\et$, and $\ez$ directions, respectively. {The remaining boundary condition \eqref{sd14} involves the slope of the edge in the $\er$-direction and, thus, expresses moment balance on $\calC$}. 

Up to the order considered, the partial-differential equation \eqref{sd10} imposes no restrictions on the in-plane displacements $u$ and $v$. Hence, $w$ and the in-plane components $u$ and $v$ are coupled only on the boundary of $\calR$. Also, $u$ and $v$ are absent from the boundary conditions \eqref{sd13} and \eqref{sd14}. Thus, \eqref{sd10} and the associated boundary conditions \eqref{sd13} and \eqref{sd14} may be used to determine $w$, independently. Satisfaction of \eqref{sd10} and \eqref{sd11}--\eqref{sd14} at the trivial solution ($u=v=w=0$) results is a relation,
\be
\iota=\half-\nu,
\label{sd144}
\ee
for the Lagrange multiplier $\lambda$ which is analogous to a result obtained by Giomi \& Mahadevan~\cite{gm} in their work on soap films bound by inextensible, elastic filaments. Next, using \eqref{sd144} in \eqref{sd12} yields
\be
\iota=\text{constant.}
\label{sol31}
\ee
The $\et$-component of force balance on the boundary of a discoidal HDL particle therefore requires that the Lagrange multiplier $\iota$ be uniform.

\section{Solving the system of equations}
\label{see}

{Along with conditions \eqref{sd144} and \eqref{sol31}, the partial-differential equation \eqref{sd10} and boundary conditions \eqref{sd13} and \eqref{sd14},
suffice to completely determine the transverse displacement $w$.} In addition, using \eqref{sd144} in \eqref{sd11} yields
\begin{multline}
\big[u+2u_{\theta\theta}+u_{\theta\theta\theta\theta}+\nu(u_{\theta\theta}+u)
\\
-\epsilon (u_{\theta\theta}+2u_{\theta\theta\theta\theta}+u_{\theta\theta\theta\theta\theta\theta})\big]_{\rho=1}=0,
\label{sol4}
\end{multline} 
which ensures the in-plane balance of forces at the boundary {and} should be accompanied by \eqref{sdinex} (or an equivalent integrated version thereof).

\subsection{In-plane deformation}
\label{id}

Equation \eqref{sol4} is an ordinary-differential equation with constant coefficients. In view of the periodicity of $u$ (i.e., $u{({1},\theta)}=u{(1,}\theta+2\pi k)$,  $\forall k\in \mathbb{Z}$), \eqref{sol4} admits a representation of the form
\be
u{({1},\theta)}=U\sin(m\theta)\quad(m\in\mathbb{Z}).
\label{sol6}
\ee
Substitution of \eqref{sol6} in \eqref{sol4} yields a characteristic equation
\be
(m^2-1) \big[ (m^2-1) +m^2(m^2-1) \epsilon-\nu \big]=0.
\label{sol7}
\ee
One solution of \eqref{sol7} is $m^2=1$, which corresponds to the planar rigid body translation and is of no physical interest. Otherwise, \eqref{sol7} yields a critical value, 
\be
\nu^{\rm i}_m=m^2-1+m^2(m^2-1)\epsilon,
\label{sol8}
\ee
$\nu$ for each planar mode $m$. Granted that $\epsilon \geq 0$, the lowest critical value of $\nu$ corresponds to $m=2$ and is given by
\be
\nu^{\rm i}_{\rm c}=\nu^{\rm i}_2=3+12\epsilon.
\label{sol81}
\ee
The value $\nu^{\rm i}_{\rm c}=3$ arising for $\epsilon=0$ is consistent with the results obtained by Chen \& Fried~\cite{CF} for a circular soap film bound by an inextensible, elastic filament. 
\subsection{Transverse displacement}
\label{td}

Modulo a rigid translation, the general solution of the partial-differential equation \eqref{sd10} {is} 
\begin{multline}
w(\rho,\theta)=a_0I_0(\zeta \rho) 
+\sum\limits_{n=1}^\infty\big(c_n \rho^n+a_n I_n(\zeta \rho)\big)\cos(n \theta)
\\
+\sum\limits_{n=1}^\infty\big(d_n \rho^n+b_n I_n(\zeta \rho)\big) \sin(n \theta),
\label{sol11}
\end{multline}
where $I_i$, $i\in\mathbb{N}$, is a modified Bessel function of the first kind. Substituting \eqref{sol11} into the boundary conditions \eqref{sd13} and \eqref{sd14} and invoking \eqref{sd144} and \eqref{sol31} results in an eigenvalue problem leading to the dispersion relation 
\begin{multline}
\Big[ -\frac{\eta}{4}\big(\zeta^3 I_n'''(\zeta)+\zeta^2 I_n''(\zeta) \big)
\\
+(\frac{\eta}{4}+\nu) \zeta I_n'(\zeta)+(\frac{\eta}{4}-\bar \eta)n^2 \zeta I_n'(\zeta)+n^4 I_n(\zeta)
\\
-(\nu+1-\bar \eta+\frac{\eta}{2})n^2 I_n(\zeta) \Big] \big[ \bar \eta n (1 -n)\big]
\\
- \Big[  -\frac{\eta}{4} n (n-1)^2+(\frac{\eta}{4}+\nu) n+(\frac{\eta}{4}-\bar \eta)n^3 
\\
+n^4 -(\nu+1-\bar \eta+\frac{\eta}{2})n^2 \Big]
\\
\Big[ \frac{\eta}{4} \zeta^2 I_n''(\zeta)+(\frac{\eta}{4}+\bar \eta) \big( \zeta I_n'(\zeta) -n^2 I_n(\zeta) \big) \Big]=0.
\label{sol18}
\end{multline}
{The terms involving $c_1$ and $d_1$ in \eqref{sol11} represent rigid body rotations about the diameter of domain $\calR$ and, thus, are physically irrelevant.} In addition, the requirements $a_0=a_1=b_1=0$ must be met to satisfy the boundary conditions \eqref{sd13} and \eqref{sd14} for $n=0$ and $n=1$. Thus, $n=2$ is the first nontrivial mode of the transverse deformation $w$. Due to its complexity, \eqref{sol18} will be studied numerically and discussed in the final section. {Whereas $\nu$ is treated as a control parameter, $\eta$ and $\bar\eta$ are treated as known input parameters}. {The solution of \eqref{sol18}, which distinguishes the critical surface tension for each transverse mode $n$, is denoted by $\nu^{\rm t}_n$.} Thus, $w$ can be written as
\be
w(\rho,\theta)=\sum\limits_{n=1}^\infty \omega_n(\rho) \Theta_n(\theta),
\label{sol16}
\ee
with{
\be
\left.
\begin{split}
\omega_n(\rho)&=  I_n(\zeta \rho) + \gamma_n \rho^n,
\\[4pt]
\Theta_n(\theta)&= a_n  \cos n \theta +b_n \sin n \theta,
\end{split}
\right\}
\label{sol17}
\ee  
}and 
\be
\gamma_n=\frac{\frac{\eta}{4} \zeta^2 I''_n(\zeta)+(\frac{\eta}{4}+\bar \eta) \big( \zeta I'_n(\zeta)-n^2I_n(\zeta)\big)}{\bar \eta n(n-1)}.
\label{sol151}
\ee

\section{Stability of a flat circular HDL particle}

Stability of the equilibrium configuration can be addressed by checking the sign of the second variation $\ddot \calH$ of the functional $\calH$. Consistent with our notation for the first variation, a superposed double dot indicates the second variation. The quantity $\ddot\calH$ can be decomposed into a sum
\begin{align}
{\ddot {\calH}}={\ddot {\calH}}_{\rm i}+{\ddot {\calH}}_{\rm t},
\label{stab1}
\end{align}  
of a purely planar component
\begin{multline}
{\ddot {\calH}}_{\rm i}=\int^{2\pi}_{0} \big[ (\nu+1)\dot{u}+(\nu+2-\epsilon) \dot{u}_{\theta\theta}
\\
+(1-2\epsilon) \dot{u}_{\theta\theta\theta\theta}+\dot{u}_{\theta\theta\theta\theta\theta\theta}\big] \dot{u} \dtheta
\label{stab110}
\end{multline}  
and a purely transverse component
\begin{align}
{\ddot {\calH}}_{\rm t}=&\int^{2 \pi}_{0} \int^{1}_{0} \Big[\frac{\eta}{4} \Delta^2 \dot {w} -\nu \Delta \dot {w} \Big] \dot {w} \rho \dro \dtheta
\notag\\
&+\int^{2\pi}_{0} \Big[ \big( \nu+\frac{\eta}{4}\big) \dot {w}_\rho-\frac{\eta}{4}\dot {w}_{\rho\rho}-\frac{\eta}{4} \dot {w}_{\rho\rho\rho} 
\notag\\
&+(\nu+1-\bar \eta+\frac{\eta}{2})\dot {w}_{\theta\theta}
+\big( \bar \eta-\frac{\eta}{4}\big) \dot {w}_{\rho\theta\theta}+\dot {w}_{\theta\theta\theta\theta} \Big] \dot {w} \dtheta
\notag\\
&+\int^{2\pi}_{0} \Big[ \big( \frac{\eta}{4}+\bar \eta \big) \dot {w}_\rho +\frac{\eta}{4} \dot {w}_{\rho\rho}+\big(\frac{\eta}{4}+\bar \eta\big) \dot {w}_{\theta\theta} \Big] \dot {w}_\rho \dtheta.
\label{stab111}
\end{align}  
The decoupling of the planar and transverse displacements in \eqref{stab1} enables separate studies of the stability of a discoidal HDL particle to planar and transverse perturbations. 

Planar stability requires that 
\be
{\ddot {\calH}}_{\rm i} >0.
\label{stab2}
\ee  
Using a Fourier expansion, the in-plane variation $\dot{u}$ may be expressed as
\be
\dot {u}{({1},\theta)}=\sum\limits_{m=1}^\infty f_m \sin m\theta,
\label{stab3}
\ee
which{,} on substitution into \eqref{stab2}, yields
\begin{multline}
\sum\limits_{m=1}^\infty f^2_m (m^2-1) \big[ (m^2-1)+m^2(m^2-1)\epsilon-\nu\big] > 0.
\label{stab4}
\end{multline}  
Since the variation $\dot{u}$ is arbitrary, the coefficients $f_m$ in \eqref{stab3} are independent and each term of the summand in \eqref{stab4} must separately satisfy the inequality \eqref{stab4}. In response to planar perturbations, a flat, circular HDL particle therefore obeys
\begin{equation}
\left.
\begin{split}
& \nu < \nu^{\rm i}_m:
\quad
{\rm stable},
\\[2pt]
& \nu > \nu^{\rm i}_m:
\quad
{\rm unstable},
\end{split}
\right\}
\label{stab5}
\end{equation}  
with $\nu^{\rm i}_m$ given in \eqref{sol8}.

Transverse stability requires that
\begin{align}
{\ddot {\calH}}_{\rm t} >0.
\label{stab6}
\end{align}  
Similar to the strategy used to investigate stability with respect to planar perturbations, a general transverse variation $\dot{w}$ can be expanded in a Fourier series. However, since the coefficients of each mode in the Fourier expansion of $\dot{w}$ are independent, it is, without loss of generality, possible to consider
\be
\dot{w}(\rho{,\theta})=\chi_n(\rho)\cos(n\theta) 
\quad
(n \in \mathbb{N}),
\label{stab7}
\ee
with $\chi_n$ being an arbitrary function. Determination of conditions necessary and sufficient to ensure \eqref{stab6} for $\chi_n$ arbitrary appears to be challenging. An alternative approach invokes the Rayleigh--Ritz variational method, in which $\chi_n$ is approximated by sum of known functions multiplied by unknown coefficients. The known functions must satisfy the geometrical boundary conditions but may otherwise be chosen arbitrarily. Guided by the structure of the general solution \eqref{sol11}, consider the \emph{Ansatz}
\be
\chi_n(\rho)=g_n \rho^n+h_n I_n(\zeta \rho),
\label{stab77}
\ee
where $g_n$ and $h_n$ are independent unknown coefficients. Substitution of \eqref{stab77} in \eqref{stab7}, and subsequently in \eqref{stab6}, and evaluating the relevant integrals yields
\begin{align}
{\ddot {\calH}}_{\rm t}=\frac{\pi}{2}[V]^{\trans} [M] [V]>0,
\label{stab78}
\end{align}  
for each $n$, {with $[V]=[g_n \ h_n]^{\trans}$ and $[M]$ a $2\times2$ matrix provided in the Appendix.} The condition necessary and sufficient for \eqref{stab78} to be satisfied is that $[M]$ be positive-definite, namely that its components obey
\begin{align}
M_{11}>0,
\quad
M_{11} M_{22}>(M_{12})^2.
\label{stab79}
\end{align}  
As a consequence of \eqref{stab79}$_1$, it follows that
\be
\nu < n(n+1)-2n \bar \eta.
\label{stab81}
\ee
Due to its complexity, \eqref{stab79}$_2$ will be studied numerically and discussed in the final section. 

\subsection{Onset of instability}

The {onset of instability} corresponds to the vanishing of the second variation $\ddot{\calH}$ of $\calH$. It can be shown that the critical values $\nu^{\rm i}_n$ and $\nu^{\rm t}_n$ of the surface tension correspond, respectively, to the onset of the planar and transverse instability. For $\nu=\nu^{\rm i}_n$ and $\nu=\nu^{\rm t}_n$, the solutions \eqref{sol6} and \eqref{sol16} can be used in \eqref{stab110} and \eqref{stab111}, respectively. Specifically, {for each $m$ and $n$, $\dot{u}$ and $\dot{w}$ can be expressed as 
\be
\left.
\begin{split}
\dot {u}({1},\theta)&=\dot U \sin(m\theta),
\\[4pt]
\dot{w}(\rho,\theta)&= \omega_n(\rho)(\dot a_n  \cos n \theta +\dot b_n \sin n \theta),
\end{split}
\right\}
\label{stab333}
\ee
where $\dot U$, $\dot a_n$, and $\dot b_n$ are the variations of the coefficients $U$, $a_n$, and $b_n$.} Regarding \eqref{stab333}$_1$ and \eqref{sol4}, it is readily observed that $\ddot{\calH}_{\rm i}$ vanishes identically. Thus, \eqref{sol7} determines the planar instability requirement and delivers the critical surface tension $\nu^{\rm i}_m$ given in \eqref{sol8}. Also, in view of \eqref{stab333}$_2$ and the equilibrium conditions \eqref{sd10}, \eqref{sd13}, and \eqref{sd14}, $\ddot{\calH}_{\rm t}$ vanishes. Thus, the dispersion relation \eqref{sol18} delivers the condition necessary for the onset of transverse instability. 

The connection between the stability conditions \eqref{stab79} and the critical value $\nu^{\rm t}_n$ at the onset of instability will be discussed in the next section. 

\section{Numerical results and discussion}
\label{sec:num}
 \begin{figure} [t] 
  \centering
  \includegraphics [height=2.5in] {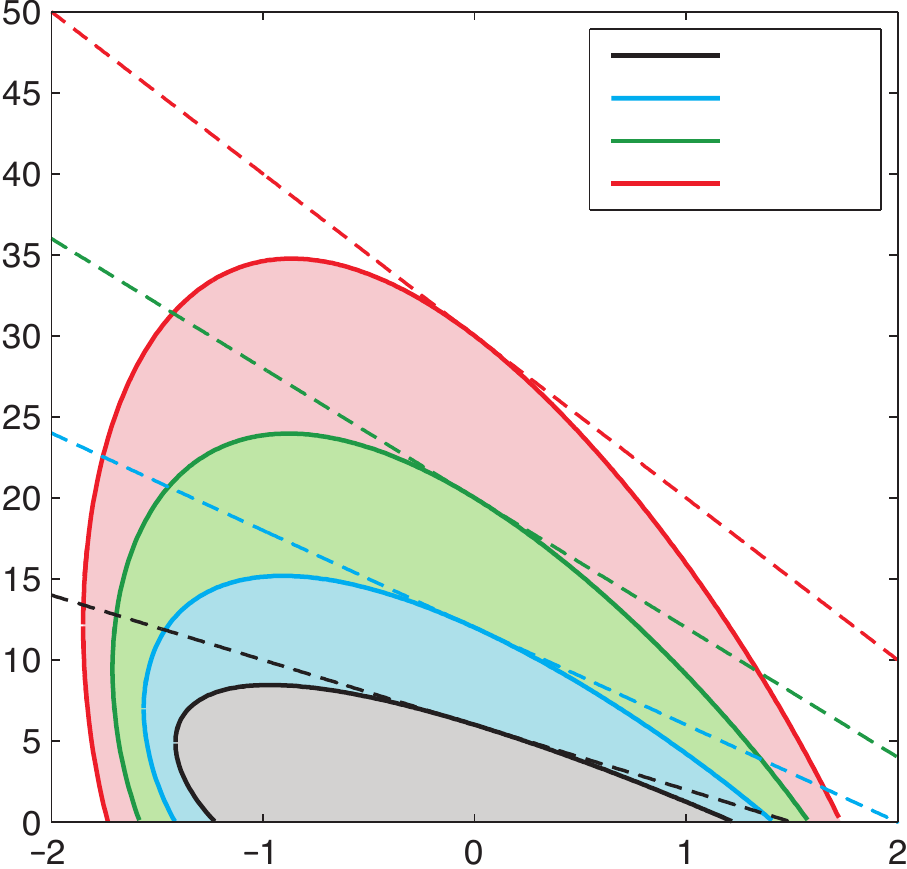}
  \put(-93,-11){$\bar \eta$}   
  \put(-199,90){\begin{sideways}$\nu$ \end{sideways}}
  \put(-35,166.5){$n=2$}
  \put(-35,158){$n=3$}
  \put(-35,149){$n=4$}
  \put(-35,140){$n=5$}
\caption{Stability plane showing the domains where a flat circular discoidal HDL particle is stable or unstable under transverse perturbations. Solid lines show $\nu^{\rm t}_n$, namely the solution of the dispersion equation \eqref{sol18}. Regions below the dashed lines are the domains where the stability requirement \eqref{stab79}$_1$ is met.}
\label{f2}
\end{figure}
 \begin{figure} [t] 
  \centering
  \includegraphics [height=2.4in] {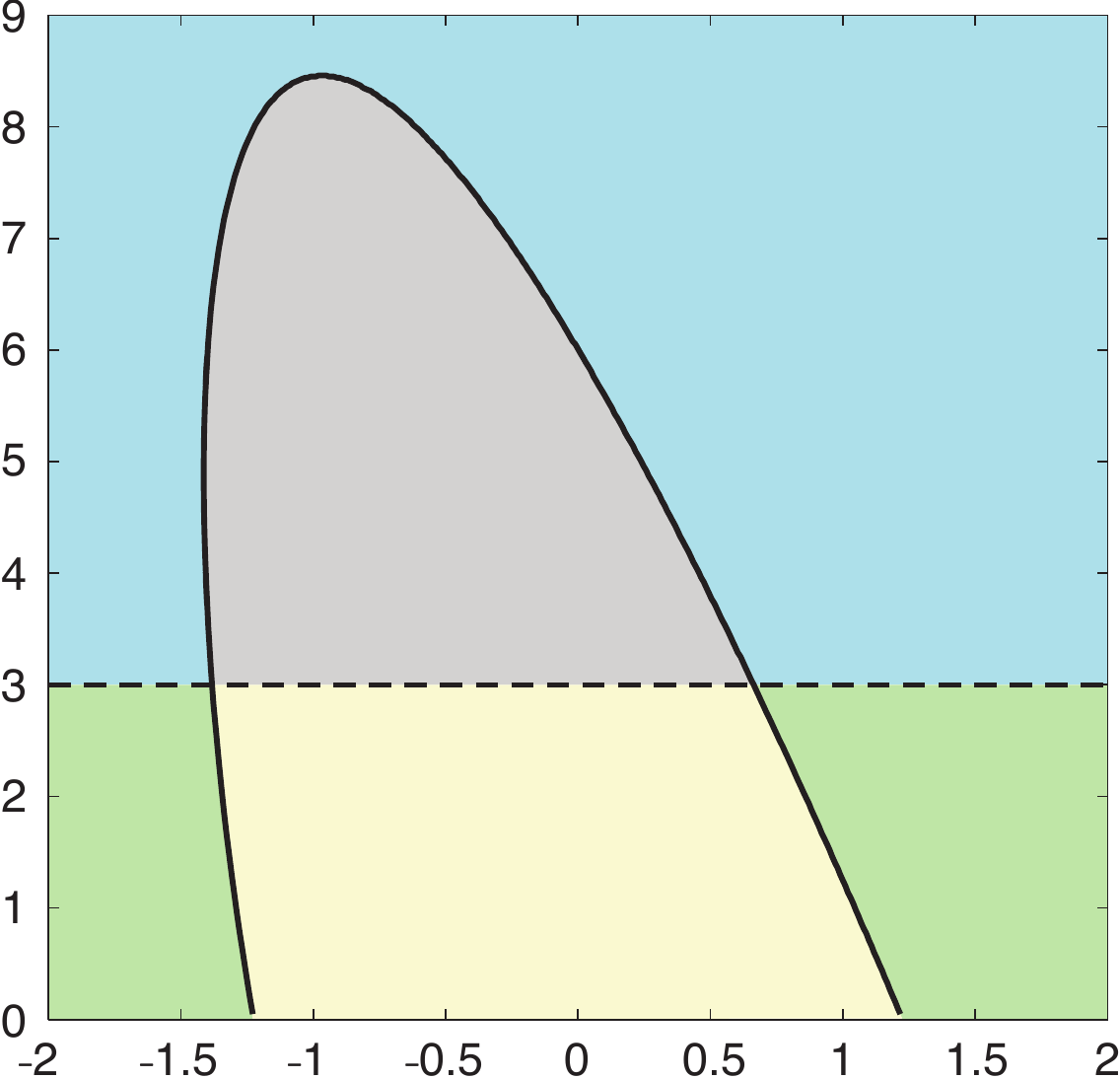}
  \put(-90,-11){$\bar \eta$}   
  \put(-190,88){\begin{sideways}$\nu$ \end{sideways}}
 \put(-103,35){$\rm(a)$}
 \put(-30,35){$\rm(b)$}
 \put(-119,100){$\rm(c)$}
 \put(-46,100){$\rm(d)$}
\caption{Stability plane of a flat circular HDL particle under transverse ($n=2$) and planar ($m=2$) perturbations, including four distinct regions (a)--(d). While the solid line shows the variation of $\nu^{\rm t}_2$, the dashed line shows {${\nu}^{\rm i}_2=3$} for $\epsilon=0$ (as provided in \eqref{sol81}).}
\label{f3}
\end{figure}
 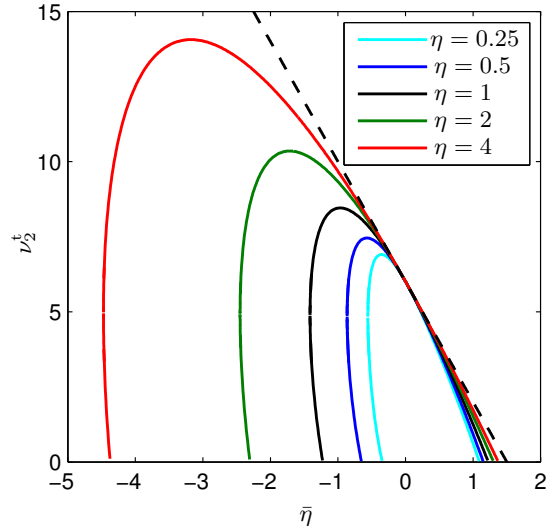
\begin{figure} [t] 
  \centering
  \includegraphics [height=2.9in] {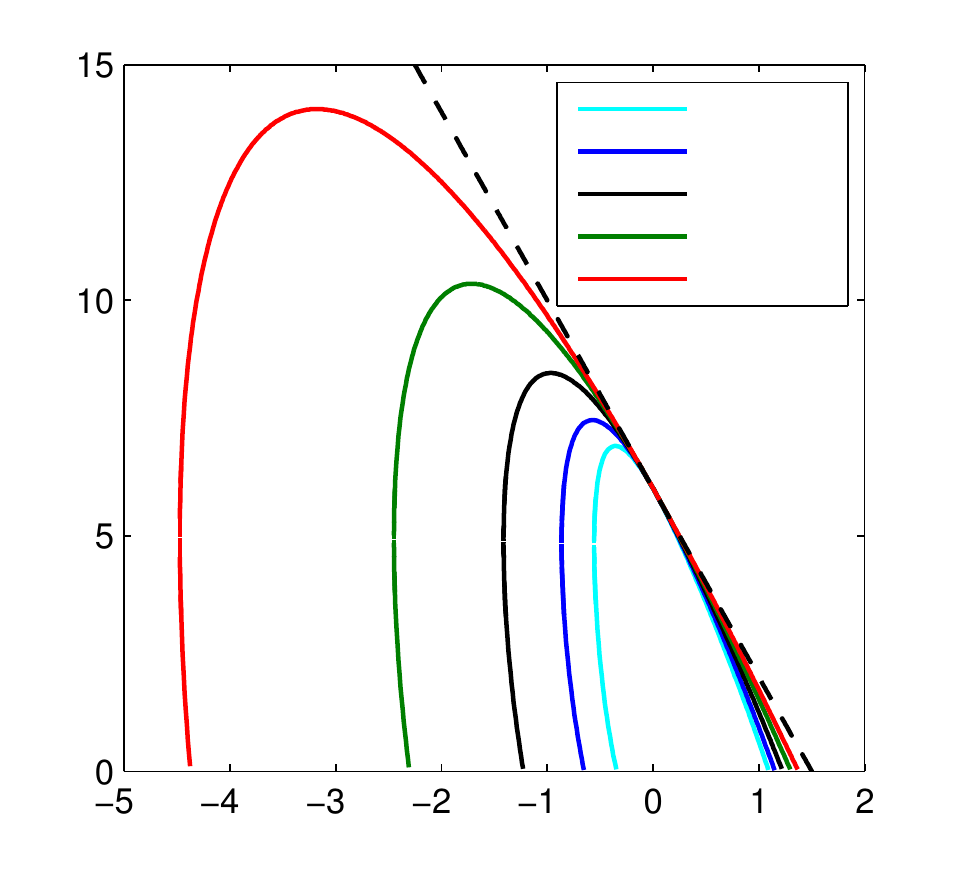}
  \put(-113,1){$\bar \eta$}   
  \put(-222,102){\begin{sideways}$\nu^{\rm t}_2$ \end{sideways}}
 \put(-63.5,181){$\eta=0.25$}
 \put(-62,170.5){$\eta=0.5$}
 \put(-62,160.5){$\eta=1$}
 \put(-62,150){$\eta=2$}
 \put(-62,140){$\eta=4$}
\caption{Effect of $\eta$ on the variation of the critical surface tension $\nu^{\rm t}_2$ with $\bar\eta$. Requirement \eqref{stab79}$_1$ is met in the region below the dashed line.}
\label{f4}
\end{figure}
Results from numerical studies based on the model are described next. Regarding the various input parameters, it seems reasonable to fix {some} of them. In particular, the splay modulus $\mu$ and the bounding loop bending stiffness $\alpha$ are kept fixed, unless mentioned otherwise. Due to the lack of data for the bending stiffness or persistence length of apoA-I, existing data for the persistence length of apolipoprotein C-II chains, which are another common component of lipoprotein particles, are used. The input parameters are merely used to illustrate the primary features of the problem; modest deviations from their exact values should not significantly affect the nature of the stability. {Hatters et al.~\cite{Hatters} report that the persistence length of apolipoprotein C-II is approximately 36~nm}, which corresponds to a bending stiffness of (36~nm)$k_B T$, with $k_B$ {Boltzmann's constant and $T$ the absolute temperature}. Assuming a double-belt apolipoprotein structure for the bounding loop yields $\alpha\sim(70\text{~nm})k_B T$. A {representative} value $\mu\approx0.5\times10^{-19}$ is used for the splay modulus of a lipid bilayer~\cite{Marsh} and, following \cite{Miyazaki3,Atkinson,Blanche,Brouillette2,Barter,Li}, assume that the reference HDL particle has diameter $2R \approx10$~nm. With these choices, \eqref{nd2}$_2$ yields $\eta\approx0.83$. {It thus seems reasonable to use $\eta=1$.}

Figure~2 depicts the transverse stability of a flat circular HDL particle for different values of the surface tension $\nu$ and the saddle-splay modulus $\bar\eta$. {Only the} first four modes {are} considered. The solid lines indicate the variation of the critical surface tension $\nu^{\rm t}_n$ with $\bar\eta$ for each mode, obtained from the dispersion relation \eqref{sol18}. For various values of $\nu$ and $\bar\eta$ in the stability plane, the second variation condition \eqref{stab79} has been used to carefully determine the nature of stability in different regions of the $(\nu,\bar\eta)$-plane. Whereas the necessary condition \eqref{stab79}$_1$ limits the stable domain to the region below the {dashed} lines, \eqref{stab79}$_2$ limits the stable domain exactly into the region enclosed by each solid line. The intersection of \eqref{stab79}$_1$ and \eqref{stab79}$_2$ determines the shaded region enclosed by each solid line as the domain where a flat circular HDL particle is stable under a transverse perturbation with mode $n$. Outside each enclosed region, the particle is unstable under a perturbation with mode $n$. {Evidently, the solid lines correspond to the onset of instability, namely the point at which an exchange of stability occurs.} Interestingly, {for each $n$, the stable region for mode $n$} is contained in the stable region of mode $n+1$. It is therefore evident that within the stable region for $n=2$ the particle is stable with respect to all higher modes. {It is also found that, within the stable region enclosed by each solid line, stability is enhanced by negative values of the saddle-splay modulus $\bar\eta$}, as the stable domain for $\bar\eta<0$ is larger than that for $\bar\eta>0$. {Finally, it is noteworthy that, while the dispersion equation \eqref{sol18} has two roots for sufficiently large negative $\bar\eta$, it otherwise has only one root.} 

{Figure~3 depicts the stability plane of a flat circular HDL particle under transverse (saddle-like) and planar (elliptical) perturbations. While the solid line corresponds to $\nu^{\rm t}_2$, the dashed line corresponds to ${\nu}^{\rm i}_2=3$ for $\epsilon=0$ (given in \eqref{sol81}).} For other values of $\epsilon\geq0$, the dashed line is merely shifted upward while remaining straight and horizontal. According to \eqref{stab5}, in the region below the dashed line, a planar discoidal HDL particle is stable. The intersection of the transverse and planar stable and unstable regions determines four distinct regions. In region $\rm(a)$, a discoidal HDL particle is stable under both transverse and planar perturbations. {Thus, a flat, circular particle should be observable only for values of $\nu$ and $\bar{\eta}$ in region $\rm{(a)}$. In region $\rm(b)$, a discoidal HDL particle is stable under planar perturbations but is destabilized by transverse saddle-like perturbations. In region $\rm(c)$, a discoidal HDL particle is stable under transverse perturbations but is unstable to planar perturbations. Thus, for values of $\nu$ and $\bar{\eta}$ in region $\rm (c)$, a noncircular flat HDL particle should be observed. Lastly, in region $\rm(d)$, a discoidal HDL particle is unstable under both transverse and planar perturbations. Hence, for values of $\nu$ and $\bar{\eta}$ in region $\rm{(d)}$, neither circular nor flat HDL particles are observable.} 

{So far, it has been assumed that the dimensionless parameter $\eta$ is fixed while allowing the other dimensionless parameters $\nu$ and $\bar\eta$ to vary. Regarding \eqref{nd2}$_2$, if the radius $R$ is held fixed, the constancy of $\eta$ requires that the ratio $\mu/\alpha$ of the splay modulus $\mu$ of the lipid bilayer and the bending rigidity $\alpha$ of the apoA-I chain to be constant. However, to have a more complete picture of the results, considering different values of $\eta$ reveals the influence of $\mu$ or $\alpha$ on the stability of discoidal HDL particles. Particularly, the effect of $\alpha$, due to lack of information on the bending modulus of apoA-I, seems essential. The variation of the critical surface tension ${\nu}^{\rm t}_2$ with $\bar\eta$ has been obtained for different values of $\eta$ and is plotted in Fig.~4. On increasing $\eta$, the region confined between each curve and the horizontal axis is magnified and extends toward more negative values of $\bar\eta$. For larger values of $\mu$ or smaller values of $\alpha$, the domain of stability for a discoidal HDL particle therefore grows.}

\vfil\eject

\section{Concluding remarks}
MD simulations of Catte et al.~\cite{catte06} reveal that gradually removing lipid molecules from discoidal HDL particles induces a transition from planar circular to nonplanar saddle-like configurations. Since the length of the apoA-I chain does not change during the depletion of lipid molecules from the bilayer of an HDL particle, {decreasing the number of lipid molecules while keeping the surface area of HDL particle fixed should increase the average spacing between neighboring lipid molecules and, hence, the} tension on the surface of particle. This is analogous to increasing the distance between the lipid molecules in each leaflet of {the} bilayer by imposing an areal stretch. It is evident from the results of Figs.~2 and 3 that increasing the surface tension $\nu$ diminishes the range of stable values for the saddle-splay modulus $\bar{\eta}$ and favors instability. {For} $\nu > \nu^{\rm i}_{\rm c}$, with $\nu^{\rm i}_{\rm c}$ given in \eqref{sol81}, a flat circular HDL particle looses its shape under in-plane perturbations. Similarly, for values of the surface tension $\nu > \nu^{\rm t}_n$, a flat circular HDL particle becomes unstable to transverse perturbations. {To reiterate}, the first planar and nonplanar unstable modes correspond respectively to planar elliptical and nonplanar saddle-like shapes. Although the linear analysis performed here is incapable of specifying the final shape that a discoidal HDL particle {might adopt}, our results, the MD simulations of Catte et al.~\cite{catte06} and Shih et al.~\cite{shih07a,shih07b}, and the experimental observations of Silva et al.~\cite{silva05}, Miyazaki et al.~\cite{miyazaki09}, Huang et al.~\cite{huang11}, and Skar-Gislinge et al.~\cite{Skar-Gislinge}, suggest that the observed planar elliptical and nonplanar saddle-like shapes of discoidal HDL particles might represent stabilized configurations of initially-flat circular particles which have become unstable due to identical types of perturbation---i.e., the planar elliptic (mode $m=2$) and nonplanar saddle-like (mode $n=2$). {This hypothesis is based on a longstanding tradition of analogous observations in structural mechanics, a tradition wherein  linearized stability analysis predicts the critical or buckling conditions under which a structure adopts a nontrivial configuration, usually a configuration with the same mode shape of the driving perturbation. The linearized analysis presented in this paper determines conditions necessary for instability of a flat circular HDL particle. Nevertheless, a comprehensive understanding of the equilibrium and stability of discoidal HDL particles requires a nonlinear analysis capable of determining nontrivial configurations involving large distortions~\cite{MF}.}

\vfil\eject

\section{Appendix}
The components of the matrix $M$ in \eqref{stab78} are
\begin{align}
&{M_{11}}= 2 n(n-1)\big(-2n \bar \eta-\nu+n(n+1)\big),
\notag\\[4pt]
&{M_{12}}={M_{21}}=(\eta/4) \big( \zeta I_n'(\zeta) (n^2+n+1)
\notag\\[2pt]
& \ \ \ \ \ \ -I_n(\zeta) n^2(n+2)+\zeta^2 I_n''(\zeta)(n-1)-\zeta^3 I_n'''(\zeta)  \big)
\notag\\[2pt]
& \ \ \ \ \ \ +\bar \eta (1-n) \big(2n\zeta I_n'(\zeta) +2n^2 I_n(\zeta) \big)
\notag\\[2pt]
& \ \ \ \ \ \ +\nu \big( \zeta I_n'(\zeta)+n I_n(\zeta) (1-2n)\big)+2 n^2 (n^2-1) I_n(\zeta),
\notag\\[4pt]
&{M_{22}}=(\eta/4) \big( 2 \zeta^2 (I_n'(\zeta))^2+2 \zeta^3 I_n'(\zeta) I_n''(\zeta)+2\zeta I_n'(\zeta) I_n(\zeta)
\notag\\[2pt]
& \ \ \ \ \ \ -4 n^2 I^2_n(\zeta)-2 \zeta^2 I_n''(\zeta) I_n(\zeta)-2 \zeta^3 I_n'''(\zeta) I_n(\zeta)\big)
\notag\\[2pt]
& \ \ \ \ \ \ +\bar \eta \big (2\zeta^2 (I_n'(\zeta))^2-4n^2 \zeta I_n(\zeta) I_n'(\zeta)+2 n^2 I^2_n(\zeta) \big)
\notag\\[2pt]
& \ \ \ \ \ \ +\nu \big( 2 \zeta I_n'(\zeta) I_n(\zeta)-2n^2 I^2_n(\zeta)\big)+2n^2(n^2-1)I^2_n(\zeta).
\label{appcoef2}
\end{align}

\end{document}